\documentclass[journal]{IEEEtran}
\IEEEoverridecommandlockouts
\usepackage[cmex10]{amsmath}
\usepackage{amssymb,bm,amsthm,mathtools}
\usepackage{graphicx}
\usepackage[caption=false,font=footnotesize]{subfig}
\usepackage{pgfplots}
\pgfplotsset{compat=1.18}
\usepackage[dvipsnames]{xcolor}
\usepackage{microtype,booktabs,cite,url}
\usepackage{algorithm,algorithmic}
\usepackage{tikz}
\usepackage{titlesec}
\usepackage{needspace}
\titlespacing*{\section}
  {0pt}{1.0ex plus 0.2ex minus 0.2ex}{0.5ex}
\titlespacing*{\subsection}
  {0pt}{0.8ex plus 0.2ex minus 0.2ex}{0.4ex}
\DeclareMathOperator{\tr}{tr}
\DeclareMathOperator{\rank}{rank}
\DeclareMathOperator{\diag}{diag}
\renewcommand{\Re}{\operatorname{Re}}
\newcommand{\Hmat}{\bm{H}}
\newcommand{\Wmat}{\bm{W}}
\newcommand{\hk}{\bm{h}_k}
\newcommand{\wk}{\bm{w}_k}
\newcommand{\CN}{\mathcal{CN}}
\newcommand{\CC}{\mathbb{C}}
\newcommand{\bPhi}{\bm{\Phi}}
\newcommand{\bOmega}{\bm{\Omega}_\perp}
\newcommand{\zmat}{\bm{Z}}
\newcommand{\bLambda}{\bm{\Lambda}}
\newcommand{\bP}{\bm{P}}
\newtheorem{theorem}{Theorem}
\newtheorem{lemma}{Lemma}

\newtheoremstyle{remarkstyle}{1.5pt}{1.5pt}{\normalfont}{0pt}{\itshape}{.}{.5em}{}
\theoremstyle{remarkstyle}

\newtheoremstyle{propositionstyle}{1.5pt}{1.5pt}{\normalfont}{0pt}{\itshape}{.}{.5em}{}
\theoremstyle{propositionstyle}
\newtheorem{proposition}{Proposition}
\newtheoremstyle{corollarystyle}{1.5pt}{1.5pt}{\normalfont}{0pt}{\itshape}{.}{.5em}{}
\theoremstyle{corollarystyle}
\newtheorem{corollary}{Corollary}
\makeatletter
\renewenvironment{proof}[1][\proofname]{\par
  \pushQED{\qed}\normalfont\topsep2\p@\@plus1\p@\relax
  \trivlist\item[\hskip\labelsep\itshape #1\@addpunct{.}]\ignorespaces
}{\popQED\endtrivlist\@endpefalse}
\def\thm@space@setup{\thm@preskip=2\p@ \thm@postskip=2\p@}
\makeatother
\pgfplotsset{every axis/.style={
    label style      = {font=\footnotesize},
    tick label style = {font=\footnotesize},
    legend style     = {font=\scriptsize,draw=gray!40,fill=white,
                        fill opacity=0.9},
    grid             = both,
    grid style       = {line width=0.3pt,color=gray!25},
    major grid style = {line width=0.5pt,color=gray!45},
    minor tick num   = 1,
}}
\definecolor{colRZF}  {RGB}{161, 33, 33}
\definecolor{colWMMSE}{RGB}{ 80, 80, 80}
\definecolor{colGIM}  {RGB}{ 27,158,119}
\definecolor{colSOCP} {RGB}{ 33, 68,161}

\newcommand{\NgainFone}{2.7}    
   
\newcommand{\NGAINlo}{1.9}      
\newcommand{\NGAINhi}{3.7}     
\newcommand{\NsocpM}{0.03}      
\newcommand{\NfeasZero}{22}     
\newcommand{\NfeasGIMhi}{15}    
\begin{document}
\title{Beyond the Pseudo-Inverse: GIM Precoding for
       Per-Antenna Power Minimisation in MU-MIMO}
\author{%
  \IEEEauthorblockN{Hernan X.~Cordova}\\
  \IEEEauthorblockA{Eindhoven University of Technology,
    Eindhoven, The Netherlands\\
    h.x.cordova@tue.nl}%
  \thanks{Preprint (draft), \today. Author's version, circulated for
    dissemination; not a peer-reviewed version of record.}%
}
\markboth{Preprint --- draft}{Preprint --- draft}
\maketitle
\begin{abstract}
We study linear precoder design for multi-user MIMO downlink under
per-antenna power constraints~(PAPC), which arise naturally when each
antenna port has its own power amplifier. While the Moore--Penrose
pseudo-inverse is the unique optimal generalized inverse for
zero-forcing precoding under a total power constraint~\cite{wiesel2008}, we prove this
optimality does not extend to the MMSE/RZF case under PAPC. Any precoder
in the Generalized Inverse Matrix~(GIM) family shares the same effective
channel and per-user SINRs as the pseudo-inverse baseline, while its null-space
component reshapes the per-antenna power profile.
This separation principle enables minimising peak
per-antenna power at fixed SINR targets with zero SINR penalty. We characterise the
convex peak-power problem over this family and its active-set optimality,
establish MRT and ZF asymptotic limits, and provide a convergent first-order subgradient
algorithm. Simulations with 8--64 antennas confirm gains of
$\NGAINlo$--$\NGAINhi$\,dB over the baseline RZF, with GIM-MMSE
eliminating essentially all PAPC violations up to $20\%$ per-antenna budget
imbalance. The framework targets fully-digital massive MIMO, with cell-free and
hybrid analog-digital arrays as extensions.
\end{abstract}
\section{Introduction}
\label{sec:intro}
Linear precoding for massive MIMO downlink seeks beamforming
matrices $\Wmat\in\CC^{M\times K}$~\cite{bjornson2014optimal}.
The regularized zero-forcing / minimum mean-square-error (RZF/MMSE) precoder is
the de~facto linear baseline~\cite{bjornson2014optimal,bjornson2017massive}.
Under a \emph{total power constraint}~(TPC), authors in~\cite{wiesel2008} proved the Moore--Penrose
pseudo-inverse (PI) is the unique optimal Generalized Inverse Matrix~(GIM) for ZF precoding,
because any null-space addition increases the total Frobenius
norm without changing the effective channel.
Under \emph{per-antenna power constraints}~(PAPC), the
diagonal entries $[\Wmat\Wmat^H]_{mm}$ must individually
satisfy limits $P_m$ imposed by the power amplifier at each
port~\cite{yu2007transmitter}.
Unlike TPC, PAPC is not invariant to null-space additions,
so the PI is suboptimal in general.
PAPC arises in hybrid analog-digital
arrays~\cite{ahmed2018survey}, cell-free massive
MIMO~\cite{bjornson2020scalable}, and large aperture arrays, motivating
rigorous characterisation of its optimal GIM structure.
\begin{table}[t]
\scriptsize
\centering
\refstepcounter{table}\label{tab:notation}%
{\footnotesize\scshape Table~\thetable: Principal Notation}\par
\vspace{1.5pt}
\renewcommand{\arraystretch}{0.65}
\setlength{\tabcolsep}{1.25pt}
\begin{tabular}{@{}lp{4.35cm}@{}}
\toprule
\textbf{Symbol} & \textbf{Meaning} \\
\midrule
$\Hmat\in\CC^{K\times M}$
  & Channel matrix; $\rank(\Hmat)=K$ a.s. \\
$\Wmat\in\CC^{M\times K}$
  & Precoding matrix \\
$\sigma^2$
  & Noise variance (per receiver) \\
$\bP = \diag\{p_k\}$, $\bP^{1/2}=\diag\{\sqrt{p_k}\}$
  & User power and amplitude matrices \\
$\Hmat^\dagger = \Hmat^H(\Hmat\Hmat^H)^{-1}$
  & Right pseudo-inverse of $\Hmat$ \\
$\bOmega = \bm{I}_M - \Hmat^\dagger\Hmat$
  & Projector onto $\ker(\Hmat)$ \\
$\bPhi\in\CC^{M\times K}$
  & Null-space free parameter \\
$\bLambda = \diag\{\lambda_k\}$
  & SINR (KKT) multiplier matrix \\
$\zmat = \bm{I}_M + \tfrac{1}{\sigma^2}\Hmat^H\bLambda\Hmat$
  & Regularised Gram matrix \\
$\Wmat(\bPhi) = \Wmat_0 + \bOmega\bPhi$, $\bm{\Xi}{=}\bOmega\bPhi$
  & GIM family; correction \\
$f_m(\bPhi)=[\Wmat(\bPhi)\Wmat(\bPhi)^H]_{mm}$
  & Per-antenna power; $f=\max_m f_m$ \\
$\gamma_k(\Wmat),\;\bar{\gamma}_k$; $P_m$
  & SINR, SINR target; power budget \\
$\tau^*=\min_{\bPhi}\max_m f_m(\bPhi)$
  & Optimal (GIM) peak power \\
$\mathcal{A}{=}\{m:f_m(\bPhi^*){=}\tau^*\}$
  & Active set; baseline peak $m^\star{=}\arg\max_m f_m(\bm0)$; iterate peak $m_s$ \\
$\bm{N}\in\CC^{M\times(M-K)}$, $\bm{Y}$
  & Orthonormal $\ker(\Hmat)$ basis; coords (Alg.~\ref{alg:gim}) \\
$c^2$ (see~\eqref{eq:common_scale})
  & Common SINR-restoring scale \\
\bottomrule
\end{tabular}
\end{table}

\noindent
\textbf{Related work and terminology.} Wiesel~\emph{et~al.}~\cite{wiesel2008}
showed the ZF pseudo-inverse can be suboptimal under PAPC and optimised
throughput/fairness \emph{within the zero-interference ZF class}; those
solutions do not carry to our \emph{dual} problem---minimise peak power at fixed
SINR---on an \emph{interference-retaining} RZF baseline, where the cross term
and common scale~\eqref{eq:common_scale} differ. The null-space freedom has
since served \emph{ZF} baselines for throughput~\cite{kaviani2013optimal} or to
curb power variance/PAPR~\cite{aghdam2021lowspatial,suzuki2018papr}, while
duality~\cite{yu2007transmitter} targets \emph{average} power. Notably,
\cite{kaviani2013optimal} searches the null space because its throughput
objective \emph{moves} with that choice, whereas here the RZF/MMSE SINRs are
\emph{invariant} to it (Lemma~\ref{lem:sep}): the freedom is spent entirely on
the peak, and the lowest achievable worst-antenna power for this family had not
been characterised before---which is the gap this letter fills. Throughout, ``baseline RZF'' is the $\bPhi{=}0$ (minimum-norm) GIM
member---the RZF analogue of the ZF pseudo-inverse, not the literal
$\Hmat^\dagger$---and ``under PAPC'' refers to problem~\eqref{eq:P1}. Optimising
the RZF regularisation~\cite{zhang2025ris} is orthogonal and composes with our
shaping.

\textbf{Contributions.}
\begin{enumerate}
\item \textbf{Zero-cost power reshaping} (via the separation
  Lemma~\ref{lem:sep}): for the MMSE/RZF baseline and any $K$, the null-space
  parameter $\bPhi$ leaves every user SINR invariant while reshaping the
  per-antenna power profile---the separation that enables PAPC shaping at no
  rate cost.
\item \textbf{Prop.~\ref{prop:rzf_papc}}: under PAPC a nonzero $\bPhi$
  generically lowers the peak power, hence the baseline ($\bPhi{=}0$) RZF---the
  minimum-norm member, optimal only under a total power constraint---is strictly
  suboptimal under PAPC.
\item \textbf{Thm.~\ref{thm:equalization}}: the peak-power problem is convex
  with active-set optimality; a strictly improving GIM solution exists a.s.,
  with full equalization when the null space admits it.
\item \textbf{Cor.~\ref{cor:feasibility}}: this reduction is attained and
  grows with $M{-}K$ (empirically, Fig.~\ref{fig:fig3_gain_vs_M}); RZF
  violates PAPC in ${\sim}\NfeasZero\%$ of realisations at equal budgets.
\item \textbf{Asymptotic limits} (Props.~\ref{prop:mrt},\ref{prop:zf}):
  GIM-MMSE inherits the correct maximum-ratio transmission (MRT) limit
  ($\sigma^2{\to}\infty$) and zero-forcing (ZF) limit ($\sigma^2{\to}0$).
\item \textbf{Algorithm~\ref{alg:gim}}: a convergent single-loop subgradient
  solver tracking a paired second-order cone program (SOCP) benchmark
  to ${\sim}0.7\%$.
\end{enumerate}

\textbf{Notation.} Bold uppercase/lowercase: matrices/vectors;
$^H$: conj.\ transpose; $^*$: elementwise conj.;
$^\dagger$: pseudo-inverse (PI); $[\bm{A}]_{mm}$: $(m,m)$ entry;
$\bm{e}_m$: $m$-th canonical basis vector (so $\bm{e}_m^T\bm{A}$ is
row $m$). See Table~\ref{tab:notation}.
\section{System Model}
\label{sec:model}
We consider MU-MIMO downlink with $M$ BS antennas serving $K$ single-antenna
users ($M{\geq}K$). The received signal $\bm{y}{=}\Hmat\Wmat\bm{x}{+}\bm{z}$
follows the standard model~\cite{bjornson2014optimal}, with
$\mathbb{E}[\bm{x}\bm{x}^H]{=}\bm{I}_K$, $\bm{z}{\sim}\CN(\bm{0},\sigma^2\bm{I}_K)$,
and perfect channel state information (CSI) at the BS. The channel is i.i.d.\
Rayleigh, $H_{km}{\sim}\CN(0,1)$ ($k$ indexes users/rows, $m$ antennas/columns),
so $\rank(\Hmat){=}K$ a.s.; this fixes the measure for every ``a.s.'' statement
below, which holds under any law absolutely continuous w.r.t.\ Lebesgue measure
on $\CC^{K\times M}$.
The signal-to-interference-plus-noise ratio (SINR) of user $k$ is:
\begin{equation}
  \gamma_k(\Wmat)=|\hk^H\wk|^2/(\sum_{i\neq k}|\hk^H\bm{w}_i|^2+\sigma^2),
  \label{eq:SINR_user_k}
\end{equation}
where $\hk^H$ is the $k$-th row of $\Hmat$ (so $\hk\in\CC^{M\times 1}$) and
$\wk\in\CC^{M\times 1}$ is the $k$-th column of $\Wmat$. Each transmit antenna $m$ satisfies the individual power budget
$[\Wmat\Wmat^H]_{mm}{\leq} P_m$. The per-antenna budgets $\{P_m\}$ are distinct from
the user power matrix $\bP{=}\diag\{p_k\}$. We seek the minimax
problem:
\begin{equation}
(\mathrm{P1}):\;\min_{\Wmat}\;\max_m\;[\Wmat\Wmat^H]_{mm}
\;\;\text{s.t.}\;\;\gamma_k(\Wmat)\geq\bar{\gamma}_k,
\;\forall k.
\label{eq:P1}
\end{equation}
Problem~\eqref{eq:P1} minimises the \emph{peak} per-antenna power at fixed SINR
and is the natural route to PAPC feasibility: with common budget $P$, the
targets are feasible \emph{if} the minimal peak over the GIM family
(parametrised in Sec.~\ref{sec:gim}), $\tau^*$, obeys $\tau^*{\leq}P$---$\tau^*$
being the GIM-family optimum at the RZF operating point, an upper bound on the
true optimal peak attained by our solver to ${\sim}0.7\%$. For
\emph{heterogeneous} budgets the governing quantity is the budget-normalised
peak $\rho^*{=}\min_{\Wmat}\max_m[\Wmat\Wmat^H]_{mm}/P_m$, feasible iff
$\rho^*{\leq}1$. All schemes are SINR-scaled, so ``infeasibility'' below is a
budget violation, not an SINR outage.
\section{GIM Structure and the Separation Principle}
\label{sec:gim}
\noindent\emph{Generalized inverse family.}
For full row-rank $\Hmat$, every right inverse takes the form
$\Hmat^{\rm inv}{=}\Hmat^\dagger{+}\bOmega\bPhi$ for arbitrary
$\bPhi{\in}\CC^{M\times K}$~\cite{wiesel2008},
where $\Hmat^\dagger\!=\!\Hmat^H(\Hmat\Hmat^H)^{-1}$ minimises
the Frobenius norm and $\bOmega\!=\!\bm{I}_M\!-\!\Hmat^\dagger\Hmat$
projects onto $\ker(\Hmat)$. Since $\Hmat\bOmega{=}\bm{0}$, every member satisfies
$\Hmat(\Hmat^\dagger{+}\bOmega\bPhi){=}\bm{I}_K$.
\subsection{Separation Principle}
\label{ssec:sep}
For any baseline $\Wmat_0{\in}\CC^{M\times K}$ with $\Hmat\Wmat_0$
invertible, define the GIM precoder family:
\begin{equation}
\Wmat(\bPhi) = \Wmat_0 + \bOmega\bPhi,\qquad
\bm{\Xi}\triangleq\bOmega\bPhi,
\label{eq:gim_W}
\end{equation}
where $\bm{\Xi}{\in}\ker(\Hmat)$ is the \emph{physical correction}. We set
$\Wmat_0{=}\Wmat_{\rm RZF}$ in Sec.~\ref{sec:opt}; the theory holds for any valid
baseline.
\begin{lemma}[Separation principle]
\label{lem:sep}
For all $\bPhi{\in}\CC^{M\times K}$ the effective channel
$\Hmat\Wmat(\bPhi){=}\Hmat\Wmat_0$ is independent of $\bPhi$; hence so are all
user SINRs.
\end{lemma}
\begin{proof}
$\Hmat\Wmat(\bPhi){=}\Hmat\Wmat_0{+}\Hmat\bOmega\bPhi
{=}\Hmat\Wmat_0$, since $\Hmat\bOmega{=}\bm{0}$. As $\Hmat\Wmat(\bPhi)$ is independent of $\bPhi$, so are all user SINRs.
\end{proof}
\noindent Lemma~\ref{lem:sep} is elementary, but it is the observation on which
every result below rests: because the user SINRs do not depend on $\bPhi$, the
null-space parameter is free to reshape the per-antenna power profile at no rate
cost.
\noindent
Expanding $f_m(\bPhi){=}[\Wmat(\bPhi)\Wmat(\bPhi)^H]_{mm}$
using~\eqref{eq:gim_W} yields the per-antenna power
decomposition:
\begin{align}
f_m(\bPhi)
&= [\Wmat_0\Wmat_0^H]_{mm}
 + [\bOmega\bPhi\bPhi^H\bOmega^H]_{mm} \nonumber\\
&\quad + 2\,\Re\{[\Wmat_0\bPhi^H\bOmega^H]_{mm}\}.
\label{eq:power_decomp}
\end{align}
Under TPC the trace cross term $2\Re\{\tr\{\Wmat_0\bPhi^H\bOmega^H\}\}$ vanishes
for any baseline whose columns lie in $\mathrm{Range}(\Hmat^H)$---both the ZF
baseline $\Hmat^\dagger\bP^{1/2}$ and $\Wmat_{\rm RZF}{=}\Hmat^H\bm{Q}$ satisfy
$\bOmega\Wmat_0{=}\bm{0}$---so $\bPhi{=}\bm{0}$ minimises total power
(recovering~\cite{wiesel2008}). The \emph{diagonal} cross term
$(\bm{e}_m^T\Wmat_{\rm RZF})(\bm{e}_m^T\bOmega\bPhi)^H$, however, is nonzero
a.s.~(Prop.~\ref{prop:rzf_papc}), so $\bPhi$ can make it negative, reducing $f_m$
with no SINR penalty.
\subsection{MMSE/RZF under PAPC: Non-Trivial Null Space}
\label{ssec:nontrivial}
\begin{proposition}[PAPC: nonzero $\bPhi$ strictly reduces a unique peak]
\label{prop:rzf_papc}
Let:
\begin{equation}
    m^\star=\arg\max_m f_m(\bm 0)
    \label{eq:argmax}
\end{equation}
be the unique baseline peak antenna. If
$
\bOmega^H\bm e_{m^\star}{\neq}\bm 0,
\bm e_{m^\star}^T\Wmat_0{\neq}\bm 0,$
then there exists a nonzero null-space correction $\bPhi$ such that
$\max_m f_m(\bPhi)<\max_m f_m(\bm 0),$
while all user SINRs remain unchanged. For the RZF/MMSE baseline
under i.i.d.\ Rayleigh fading, these conditions---including uniqueness
of the baseline peak antenna $m^\star$---hold a.s.
\end{proposition}
\begin{proof}
\textbf{SINR invariance} follows directly from
Lemma~\ref{lem:sep}.
Since $f_m(\bPhi)\!=\![\Wmat(\bPhi)\Wmat(\bPhi)^H]_{mm}
\!=\!\|\bm{e}_m^T(\Wmat_0{+}\bOmega\bPhi)\|^2$
is the squared norm of an affine function of $\bPhi$,
$f_m$ is a convex quadratic in $\bPhi$.

\noindent
\textbf{Wirtinger gradient.}
By the complex matrix-differential rules
of~\cite{hjorungnes2007complex}, the Wirtinger derivative of the only
$\bPhi^H$-linear term of $f_m$,
$\tr\{\bm{e}_m^T\Wmat_0\,\bPhi^H\bOmega^H\bm{e}_m\}$, with respect to $\bPhi^*$
is read off directly as:
\begin{equation}
\nabla_{\bPhi^*}f_m\big|_{\bPhi=\bm{0}}
= \bOmega^H\bm{e}_m\bm{e}_m^T\Wmat_0.
\label{eq:gradient_at_zero}
\end{equation}

\noindent
Let $g{=}f_{m^\star}(\bm 0){-}\max_{m{\neq} m^\star}f_m(\bm 0){>}0$. As
$\bm G_{m^\star}{=}\bOmega^H\bm e_{m^\star}\bm e_{m^\star}^T\Wmat_0{\neq}\bm 0$,
the step $-t\bm G_{m^\star}$ strictly decreases $f_{m^\star}$; by continuity and
the gap $g$, for small $t{>}0$ no antenna overtakes it, so
$\max_m f_m(-t\bm G_{m^\star}){<}\max_m f_m(\bm 0)$. The hypotheses fail only on
measure-zero events ($\bm e_{m^\star}{\in}\mathrm{Range}(\Hmat^H)$,
$\bm e_{m^\star}^T\Wmat_0{=}\bm 0$, or a top-two tie), so they and the descent
hold a.s.
\end{proof}
\subsection{Null-Space Equalization of Per-Antenna Power}
\label{ssec:equalization}
\begin{theorem}[Null-space equalization and existence]
\label{thm:equalization}
Let $\Wmat(\bPhi){=}\Wmat_0{+}\bOmega\bPhi$ with
$\rank(\Hmat){=}K$ and $M{>}K$.

\noindent\textit{(i) Convexity and equalization.}
The minimax problem is convex in $\bPhi$ as:
\begin{equation}
\min_{\bPhi}\;f(\bPhi)\triangleq\max_m\;f_m(\bPhi)
\label{eq:minimax_phi}
\end{equation}
At any optimum $\bPhi^*$ (value $\tau^*$) the active set
$\mathcal{A}{=}\{m{:}f_m(\bPhi^*){=}\tau^*\}$ meets a convex
subgradient-balance condition; full equalization (all $f_m{=}\tau$) holds iff
that quadratic system is solvable---characterised in~(i.2)---and then $\bPhi^*$
equalizes every antenna to $\tau^*$.

\noindent\textit{(ii) Existence of a strictly improving solution.}
For any $\Hmat$ with $\rank(\Hmat){=}K{<}M$ such that
$\bm{e}_m{\notin}\mathrm{Range}(\Hmat^H)$ and $\bm{e}_m^T\Wmat_0{\neq}\bm{0}$ $\forall m$,
a global minimiser $\bPhi^*{\neq}\bm{0}$ exists with
$f(\bPhi^*){<}f(\bm{0})$.
These conditions hold for all $\Hmat$ outside a measure-zero
algebraic set, hence a.s.\ under the adopted i.i.d.\ Rayleigh
(Lebesgue-absolutely-continuous) law.

\noindent\textit{(iii) Uniqueness of the physical minimiser.}
If every $(M{-}K)$ rows of $\bm{N}$ are linearly independent, the physical
correction $\bm{\Xi}^*{=}\bm{N}\bm{Y}^*$ is unique and continuous in $\Wmat_0$
at fixed $\Hmat$; this row condition holds a.s.\ under the adopted law.
\end{theorem}
\begin{proof}
\textbf{(i) Convexity.}
Each $f_m(\bPhi){=}\|\bm{e}_m^T(\Wmat_0+\bOmega\bPhi)\|^2$
is the squared $\ell_2$ norm of an affine function of
$\bPhi$, hence a convex quadratic in $\bPhi$.
The pointwise maximum $f(\bPhi){=}\max_m f_m(\bPhi)$ of
finitely many convex functions is convex~\cite[Sec.~3.2]{
boyd2004convex}.

\noindent
\textbf{(i.1) Active-set balance.}
With $\mathcal{A}{=}\{m:f_m(\bPhi^*){=}\tau^*\}$, optimality gives
$\bm{0}{\in}\partial f(\bPhi^*){=}\mathrm{conv}\{\nabla f_m(\bPhi^*):m{\in}\mathcal{A}\}$
(Danskin's theorem: the subdifferential of a pointwise maximum is the convex
hull of the gradients of the active pieces). Since
$\nabla_{\bPhi^*}f_m{=}\bOmega^H\bm{e}_m\bm{e}_m^T(\Wmat_0{+}\bm{\Xi})$,
this is the projected active-gradient balance: there exist $\theta_m{\geq}0$
with $\sum_m\theta_m{=}1$ on $\mathcal{A}$ such that:
\begin{equation}
\bOmega^H\!\sum_{m}\theta_m\bm{e}_m\bm{e}_m^T(\Wmat_0{+}\bm{\Xi}^*)=\bm{0},
\label{eq:stationarity}
\end{equation}
with $\bm{\Xi}^*{=}\bOmega\bPhi^*$ being the optimal physical correction.

\noindent
\textbf{(i.2) Full equalization is conditional.}
The correction supplies $2(M{-}K)K$ real degrees of freedom, but
$f_m(\bPhi){=}\tau$ is \emph{quadratic}, so a degree count alone does not
guarantee a real solution; full equalization holds conditionally, whenever such
a $(\bPhi,\tau)$ exists---confirmed in the tested regime (Sec.~\ref{sec:sims}).

\noindent
\textbf{(ii) Existence.}
By Prop.~\ref{prop:rzf_papc}, a descent step at $\bPhi{=}\bm{0}$
strictly decreases $f$. Although $\bPhi$ is redundant along
$\ker(\bOmega)$, $f$ depends only on the physical correction
$\bm{\Xi}{=}\bOmega\bPhi{\in}\mathrm{Range}(\bOmega)$, and
$\max_m\|\bm{e}_m^T(\Wmat_0{+}\bm{\Xi})\|^2
{\geq} M^{-1}(\|\bm{\Xi}\|_F{-}\|\Wmat_0\|_F)^2$;
hence $f$ is coercive in $\bm{\Xi}$ over this closed subspace and
attains a global minimum $\bm{\Xi}^*$. With the strict
improvement above, $\bm{\Xi}^*{\neq}\bm{0}$, so $\bPhi^*{\neq}\bm{0}$.

\noindent
\textbf{(iii) Uniqueness.} In coordinates $\bm{\Xi}{=}\bm{N}\bm{Y}$, at
$\tau^*{>}0$ let $J{=}\{m{:}\theta_m{>}0\}$. If $|J|{\le}M{-}K$, row-independence
of $\bm{N}$ and~\eqref{eq:stationarity} force $\bm{e}_m^T\Wmat(\bPhi^*){=}\bm{0}$
on $J$, contradicting $\tau^*{>}0$; hence
$\bm{Q}{=}\bm{N}^H\!\diag\{\theta_m\}\bm{N}{\succ}0$ and
$f(\bPhi){\ge}\tau^*{+}\lambda_{\min}(\bm{Q})\|\bm{Y}{-}\bm{Y}^*\|_F^2$, so
$\bm{\Xi}^*$ is unique and (as the bound localises the minimiser) continuous in
$\Wmat_0$; $\tau^*{=}0$ follows from injectivity of $\bm{N}$.
\end{proof}
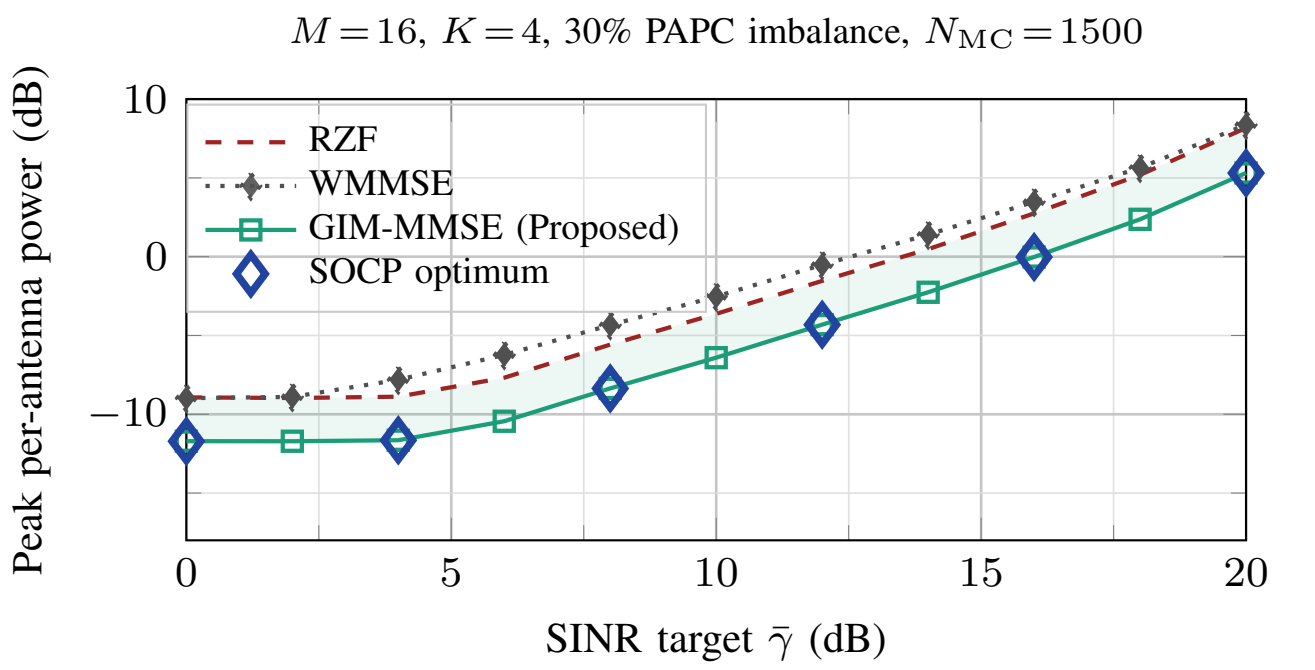
\begin{figure}[t]
\centering
\begin{tikzpicture}
\begin{axis}[
  width=0.97\columnwidth, height=4.5cm,
  xlabel={SINR target $\bar{\gamma}$ (dB)},
  ylabel={Peak per-antenna power (dB)},
  xmin=0, xmax=20, ymin=-18, ymax=10,
  legend style={font=\scriptsize, draw, fill=none,
        fill opacity=0.9, text opacity=1,
        at={(0.0,0.99)}, anchor=north west, row sep=-2pt},
  legend cell align=left,
  title={\scriptsize $M\!=\!16$, $K\!=\!4$, 30\% PAPC imbalance, $N_{\rm MC}\!=\!1500$}]

\addplot[colRZF,thick,dashed,mark=none] coordinates {
 (0,-8.928) (2,-8.969) (4,-8.885) (6,-7.68) (8,-5.579) (10,-3.627) (12,-1.535) (14,0.479) (16,2.748) (18,5.196) (20,8.17)};
\addlegendentry{RZF}

\addplot[colWMMSE,thick,dotted,mark=diamond*,mark size=2.3] coordinates {
 (0,-8.978) (2,-8.898) (4,-7.813) (6,-6.232) (8,-4.339) (10,-2.528) (12,-0.518) (14,1.4) (16,3.482) (18,5.644) (20,8.374)};
\addlegendentry{WMMSE}

\addplot[colGIM,thick,solid,mark=square,mark size=1.8] coordinates {
 (0,-11.709) (2,-11.712) (4,-11.647) (6,-10.442) (8,-8.358) (10,-6.408) (12,-4.313) (14,-2.258) (16,-0.019) (18,2.372) (20,5.312)};
\addlegendentry{GIM-MMSE (Proposed)}

\addplot[colSOCP,only marks,mark=diamond,mark size=3.6,line width=1.4pt,mark options={solid,line width=1.4pt}] coordinates {
 (0,-11.713) (4,-11.651) (8,-8.362) (12,-4.317) (16,-0.023) (20,5.308)};
\addlegendentry{SOCP optimum}

\addplot[colGIM,fill=colGIM,fill opacity=0.08,draw=none,forget plot] coordinates {
 (0,-8.928) (2,-8.969) (4,-8.885) (6,-7.68) (8,-5.579) (10,-3.627) (12,-1.535) (14,0.479) (16,2.748) (18,5.196) (20,8.17)
 (20,5.312) (18,2.372) (16,-0.019) (14,-2.258) (12,-4.313) (10,-6.408) (8,-8.358) (6,-10.442) (4,-11.647) (2,-11.712) (0,-11.709)}
 \closedcycle;
\end{axis}
\end{tikzpicture}
\caption{Peak per-antenna power vs.\ $\bar{\gamma}$ ($M{=}16$, $K{=}4$). GIM-MMSE is \emph{parallel} to RZF---a
constant $2.78$\,dB reduction at every SINR target (scale invariance)---and
coincides with the exact SOCP optimum to $<\!0.02$\,dB.}
\label{fig:fig1_power_vs_gamma}
\end{figure}

\noindent\emph{Reverse waterfilling:} $\bOmega\bPhi$ shifts power from overloaded
to underloaded ports at zero SINR cost, balancing only the \emph{binding}
antennas (Thm.~\ref{thm:equalization}); the gain rises with $M{-}K$ with
diminishing returns (Fig.~\ref{fig:fig3_gain_vs_M}).
\begin{corollary}[PAPC attainment and nested-space improvement]
\label{cor:feasibility}
The PAPC feasible set $\mathcal{F}\!=\!\{\bPhi:f_m(\bPhi){\leq} P_m,\forall m\}$
is closed and convex, and~\eqref{eq:minimax_phi} over $\mathcal{F}$ attains its
minimum when $\mathcal{F}{\neq}\emptyset$. Enlarging the correction subspace
cannot increase the optimal peak; the attainable reduction thus grows with the
null-space dimension $M{-}K$ (Fig.~\ref{fig:fig3_gain_vs_M}).
\end{corollary}
\begin{proof}
Attainment follows from continuity and coercivity in the physical
correction $\bm{\Xi}{=}\bOmega\bPhi$
(Thm.~\ref{thm:equalization}). The nested-space claim holds because
minimising the same objective over a superset cannot yield a larger
optimum.
\end{proof}
\section{Optimal GIM-MMSE Precoder}
\label{sec:opt}
\subsection{KKT-Optimal RZF/MMSE Baseline}
The KKT conditions for the SINR-constrained power-minimisation
problem, with SINR multipliers
$\lambda_k{\geq} 0$ ($\bLambda{=}\diag\{\lambda_k\}$),
yield~\cite{bjornson2014optimal}:
\begin{equation}
\wk^* {=} \sqrt{p_k}\;
\frac{\zmat^{-1}\hk}{\|\zmat^{-1}\hk\|},
\quad
\zmat {=} \bm{I}_M+\tfrac{1}{\sigma^2}\Hmat^H\bLambda\Hmat,
\label{eq:kkt_bf}
\end{equation}
where $p_k{>}0$ and the KKT-optimal $\{\lambda_k,p_k\}$ solve a coupled fixed
point. The \emph{peak} problem~\eqref{eq:P1} is different---its epigraph carries
per-antenna multipliers $\diag\{\mu_m\}$ and is \emph{not} solved
by~\eqref{eq:kkt_bf}---so we use RZF only as a baseline and shape the peak within
its GIM family. We take the representative member $\bLambda{=}\bm{I}_K$
(regularisation $\sigma^2$, so ``RZF''${=}$``MMSE''), a valid baseline for
Lemma~\ref{lem:sep} since $\Hmat\Wmat_0$ is invertible:
\begin{equation}
\Wmat_{\rm RZF} {=} \zmat^{-1}\Hmat^H\bP^{1/2}{=}\Hmat^H\bm{Q},
\quad\bP^{1/2}{=}\diag\{\sqrt{p_k}\},
\label{eq:rzf_kkt}
\end{equation}
with push-through factor
$\bm{Q}{=}(\bm{I}_K{+}\tfrac{1}{\sigma^2}\bLambda\Hmat\Hmat^H)^{-1}\bP^{1/2}$,
so its columns lie in $\mathrm{Range}(\Hmat^H)$. Applying the GIM
parameterization~\eqref{eq:gim_W},
\begin{equation}
\Wmat_{\rm GIM} = \Wmat_{\rm RZF} + \bOmega\bPhi^*,
\label{eq:gim_final}
\end{equation}
with $\bPhi^*$ from Alg.~\ref{alg:gim}. By
Lemma~\ref{lem:sep},
$\Hmat\Wmat_{\rm GIM}{=}\Hmat\Wmat_{\rm RZF}$, so all SINRs are maintained.
A common scalar $c$ restores the targets: with $S_k{=}|\hk^H\wk|^2$,
$I_k{=}\sum_{i\neq k}|\hk^H\bm{w}_i|^2$, requiring
$\gamma_k(c\Wmat){\geq}\bar{\gamma}_k$ gives:
\begin{equation}
c^2=\max\!\Big\{1,\;\max_k\frac{\bar{\gamma}_k\,\sigma^2}{S_k-\bar{\gamma}_k I_k}\Big\}
\quad(S_k{>}\bar{\gamma}_k I_k),
\label{eq:common_scale}
\end{equation}
Since $\Hmat\Wmat_{\rm GIM}{=}\Hmat\Wmat_{\rm RZF}$, the pairs $(S_k,I_k)$---hence
$c^2$---coincide for RZF and GIM, so the peak-power \emph{ratio} is invariant to
this scale.
\subsection{Asymptotic Behaviour}
Both limits are \emph{under PAPC}: the RZF direction converges (a channel
property) while the correction $\bOmega\bPhi^*$ generally persists.
\begin{proposition}[Low-SNR / large-array: convergence to MRT]
\label{prop:mrt}
As $\sigma^2{\to}\infty$, $\Wmat_{\rm GIM}$ converges to the GIM-shaped MRT
precoder, i.e.\ the GIM family built on the MRT baseline $\Hmat^H\bP^{1/2}$.
In the joint limit ($M,K{\to}\infty$, fixed $M/K$) it approaches plain MRT.
For fixed $K$ the baseline peak-to-average ratio diverges ($\Theta(\log M)$)
and the null-space correction persists (e.g., our regime test, $K{=}4$).
\end{proposition}
\begin{proof}
As $\sigma^2{\to}\infty$, $\zmat^{-1}{\to}\bm{I}_M$, so
$\Wmat_{\rm RZF}{\to}\Hmat^H\bP^{1/2}$ and $\bPhi^*{\to}\bPhi^*_{\rm MRT}$.
In the joint limit the power $\sum_k p_k|H_{km}|^2$ concentrates
(variance $O(1/K)$, so peak-to-average $\to1$), so
$\bPhi^*_{\rm MRT}{\to}\bm{0}$ and
$\Wmat_{\rm GIM}{\to}\Hmat^H\bP^{1/2}$; for fixed $K$ the per-antenna load
$\sum_k p_k|H_{km}|^2$ has an exponential tail, so its maximum over the
$M$ i.i.d.\ antennas grows as $\Theta(\log M)$ (extreme value of
sub-exponential variables), the peak-to-average diverges, and
$\bPhi^*_{\rm MRT}$ does not vanish.
\end{proof}
\begin{proposition}[High-SNR: convergence to ZF]
\label{prop:zf}
As $\sigma^2{\to}0$, $\Wmat_{\rm GIM}$ converges to a \emph{ZF} precoder
($\Hmat\Wmat_{\rm GIM}{\to}\tilde{\bP}^{1/2}{=}\sigma^2\bLambda^{-1}\bP^{1/2}$),
but \emph{not} the PI: the null-space correction persists, so
the limit is the \emph{minimum-peak-power} ZF precoder in the GIM family.
\end{proposition}
\begin{proof}
By push-through, $\Wmat_{\rm RZF}{\to}\Hmat^\dagger\tilde{\bP}^{1/2}$ as
$\sigma^2{\to}0$. Every GIM member is zero-forcing
($\Hmat\Wmat(\bPhi){=}\Hmat\Wmat_{\rm RZF}{\to}\tilde{\bP}^{1/2}$,
Lemma~\ref{lem:sep}), so the correction reshapes per-antenna power
\emph{within} the interference-free class. Baseline (in
$\mathrm{Range}(\Hmat^H)$) and correction $\bOmega\bPhi^*_{\rm ZF}$ (in
$\ker\Hmat$) are both $O(\sigma^2)$. As the ZF baseline meets
Prop.~\ref{prop:rzf_papc}, the correction is of the same order and does
not vanish relatively, so $\bPhi^*_{\rm ZF}{\neq}\bm{0}$.
\end{proof}
\subsection{Subgradient Algorithm}
The $\bPhi$-subproblem~\eqref{eq:minimax_phi} is a convex nonsmooth minimax
(Thm.~\ref{thm:equalization}(i)). Rather than an exact SOCP solve ($O(M^{3.5})$),
we run a null-basis subgradient descent (Alg.~\ref{alg:gim}), $O(M^2K)$ per step
and scalable to large $M$.
\begin{figure}[t]
\centering
\begin{tikzpicture}
\begin{axis}[
  width=0.91\columnwidth, height=4.5cm,
  xlabel={Antenna imbalance $\Delta/(P_0/M)$ (\%)},
  ylabel={Infeasibility probability},
  xmin=0, xmax=50, ymin=0, ymax=1.0,
  legend style={font=\scriptsize, draw=none, fill=none,
        fill opacity=0.9, text opacity=1,
        at={(0.0,1.05)}, anchor=north west, row sep=-2pt},
 legend cell align=left,
  title={\scriptsize $M\!=\!16$, $K\!=\!4$, $\bar{\gamma}\!=\!10$\,dB, $N_{\rm MC}\!=\!1500$}]

\addplot[colRZF,thick,dashed,mark=none] coordinates {
 (0,0.225) (5,0.233) (10,0.264) (15,0.263) (20,0.275) (25,0.316) (30,0.361) (35,0.393) (40,0.441) (45,0.519) (50,0.583)};
\addlegendentry{RZF}

\addplot[colWMMSE,thick,dotted,mark=diamond*,mark size=2.3] coordinates {
 (0,0.565) (5,0.565) (10,0.571) (15,0.603) (20,0.617) (25,0.631) (30,0.693) (35,0.705) (40,0.729) (45,0.779) (50,0.82)};
\addlegendentry{WMMSE}

\addplot[colGIM,thick,solid,mark=square,mark size=1.8,
  error bars/.cd,y dir=both,y explicit,error bar style={colGIM}] coordinates {
 (0,0.0) +- (0,0.0026) (5,0.0007) +- (0,0.0031) (10,0.002) +- (0,0.0039) (15,0.002) +- (0,0.0039) (20,0.002) +- (0,0.0039) (25,0.0033) +- (0,0.0044) (30,0.008) +- (0,0.0059) (35,0.0147) +- (0,0.0074) (40,0.0393) +- (0,0.0111) (45,0.0827) +- (0,0.015) (50,0.1547) +- (0,0.0192)};
\addlegendentry{GIM-MMSE (Proposed)}

\addplot[colSOCP,only marks,mark=diamond,mark size=3.6,line width=1.4pt,mark options={solid,line width=1.4pt}] coordinates {
 (0,0.0) (10,0.002) (20,0.002) (30,0.008) (40,0.0393) (50,0.154)};
\addlegendentry{SOCP optimum}

\addplot[colGIM,fill=colGIM,fill opacity=0.08,draw=none,forget plot] coordinates {
 (0,0.225) (5,0.233) (10,0.264) (15,0.263) (20,0.275) (25,0.316) (30,0.361) (35,0.393) (40,0.441) (45,0.519) (50,0.583)
 (50,0.155) (45,0.083) (40,0.039) (35,0.015) (30,0.008) (25,0.003) (20,0.002) (15,0.002) (10,0.002) (5,0.001) (0,0.0)}
 \closedcycle;
\end{axis}
\end{tikzpicture}
\caption{PAPC Infeasibility probability vs.\ $\Delta/(P_0/M)$. RZF violates PAPC in $22\%$ of
realisations even at equal budgets; GIM-MMSE eliminates violations up
to $20\%$ imbalance, tracking closely the SOCP feasibility5(Cor.~\ref{cor:feasibility}).}
\label{fig:fig2_feasibility}
\end{figure}
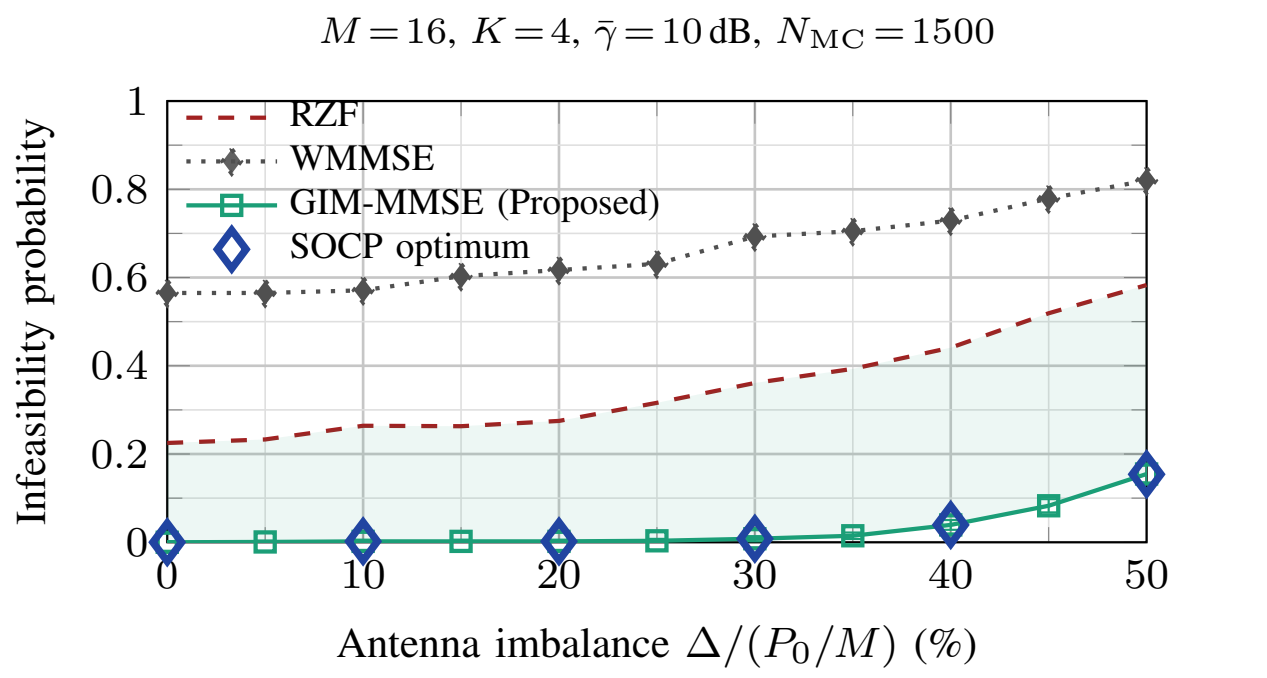

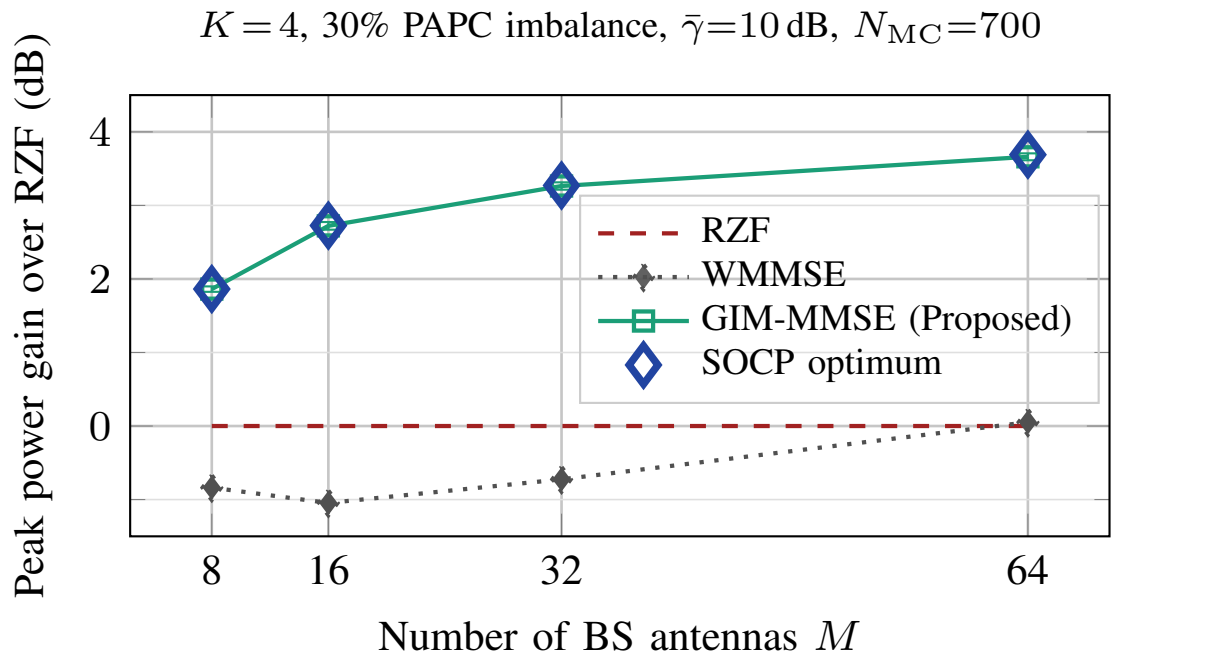
\begin{figure}[t]
\centering
\begin{tikzpicture}
\begin{axis}[
  width=0.91\columnwidth, height=4.5cm,
  xlabel={Number of BS antennas $M$},
  ylabel={Peak power gain over RZF (dB)},
  xtick={8,16,32,64}, xticklabels={8,16,32,64},
  ymin=-1.5, ymax=4.5,
  legend style={font=\scriptsize, draw, fill=none,
        fill opacity=0.9, text opacity=1,
        at={(0.99,0.30)}, anchor=south east, row sep=-2pt},
   legend cell align=left,
  title={\scriptsize $K\!=\!4$, 30\% PAPC imbalance, $\bar{\gamma}{=}10$\,dB, $N_{\rm MC}{=}700$}]

\addplot[colRZF,thick,dashed,mark=none] coordinates {(8,0)(64,0)};
\addlegendentry{RZF}

\addplot[colWMMSE,thick,dotted,mark=diamond*,mark size=2.3] coordinates {
 (8,-0.838) (16,-1.046) (32,-0.725) (64,0.052)};
\addlegendentry{WMMSE}

\addplot[colGIM,thick,solid,mark=square,mark size=1.8,
  error bars/.cd,y dir=both,y explicit,error bar style={colGIM}] coordinates {
 (8,1.862) +- (0,0.037) (16,2.722) +- (0,0.052) (32,3.264) +- (0,0.05) (64,3.661) +- (0,0.047)};
\addlegendentry{GIM-MMSE (Proposed)}

\addplot[colSOCP,only marks,mark=diamond,mark size=3.6,line width=1.4pt,mark options={solid,line width=1.4pt}] coordinates {
 (8,1.863) (16,2.726) (32,3.271) (64,3.69)};
\addlegendentry{SOCP optimum}
\end{axis}
\end{tikzpicture}
\caption{Peak-power gain over RZF vs.\ $M$. Gain grows with $M{-}K$
(reverse waterfilling; Thm.~\ref{thm:equalization}); the first-order algorithm
matches the exact SOCP optimum to within $0.03\,$dB.}
\label{fig:fig3_gain_vs_M}
\end{figure}

\noindent
The physical correction $\bOmega\bPhi{=}\bm{N}\bm{Y}$ uses orthonormal $\bm{Y}$
to remove the gauge redundancy of $\bPhi$. Step~5's $\bm{G}$ is a subgradient of
$\max_m f_m$ (Danskin,~\eqref{eq:gradient_at_zero}); its norm grows at most
linearly in $\|\bm{Y}\|$, so the square-summable schedule
($\sum_s\eta_s{=}\infty$, $\sum_s\eta_s^2{<}\infty$, met by $p{=}0.6$) keeps the
iterates bounded and best-iterate descent converges to the subproblem
optimum~\cite{nedic2001incremental,boyd2003subgradient}; a single budget
$T_\Phi{=}8000$ ($\eta_0{=}0.5$) reaches within ${\sim}0.7\%$ of the paired SOCP
for every $M$. The same solver handles the \emph{weighted} peak
$\min_{\bPhi}\max_m a_m f_m(\bPhi)$---e.g.\ $a_m{=}1/P_m$, recovering the
budget-normalised $\rho^*$ of Sec.~\ref{sec:model}---by the single
substitution $\bm{e}_m^T\Wmat\!\mapsto\!\sqrt{a_m}\,\bm{e}_m^T\Wmat$, which
stays convex quadratic; the SINR targets act through the constraints, not the
weights.
\section{Simulation Results}
\label{sec:sims}
\begin{algorithm}[t]
\caption{GIM-MMSE under PAPC}
\label{alg:gim}
\begin{algorithmic}[1]
\REQUIRE \scriptsize Channel $\Hmat$; baseline $\Wmat_{\rm RZF}$~\eqref{eq:rzf_kkt};
  orthonormal basis $\bm{N}$
  ($\bm{N}^H\bm{N}{=}\bm{I}$);
  $T_\Phi$: subgradient steps;
  step schedule $\eta_s{=}\eta_0/(s{+}1)^p$, $p\in(\tfrac12,1]$
\STATE $\bm{Y}\leftarrow\bm{0}$;\;
  $\bm{Y}^*\leftarrow\bm{0}$;\;
  $f^*\leftarrow\max_m\|\bm{e}_m^T\Wmat_{\rm RZF}\|^2$
\FOR{$s=0,\ldots,T_\Phi-1$}
  \STATE $\Wmat\leftarrow\Wmat_{\rm RZF}+\bm{N}\bm{Y}$
  \STATE $m_s\leftarrow\arg\max_m \|\bm{e}_m^T\Wmat\|^2$
    \COMMENT{\scriptsize worst-power antenna at iterate $s$}
  \STATE $\bm{G}\leftarrow(\bm{N}^H\bm{e}_{m_s})\,(\bm{e}_{m_s}^T\Wmat)$
    \COMMENT{\scriptsize subgradient of $\max_m f_m$}
  \STATE $\bm{Y}\leftarrow\bm{Y}-\eta_s\bm{G}$
  \STATE \textbf{if} $\max_m\|\bm{e}_m^T(\Wmat_{\rm RZF}+\bm{N}\bm{Y})\|^2<f^*$
    \textbf{then} $\bm{Y}^*\leftarrow\bm{Y}$, update $f^*$
    \COMMENT{\scriptsize best iterate}
\ENDFOR
\STATE $\Wmat_{\rm GIM}\leftarrow\Wmat_{\rm RZF}+\bm{N}\bm{Y}^*$
\STATE $\Wmat_{\rm GIM}\leftarrow\sqrt{c^2}\,\Wmat_{\rm GIM}$,
  $c^2$ from~\eqref{eq:common_scale}
  \COMMENT{\scriptsize restore SINR targets}
\ENSURE $\Wmat_{\rm GIM}=\sqrt{c^2}\,(\Wmat_{\rm RZF}+\bm{N}\bm{Y}^*)$
\end{algorithmic}
\end{algorithm}
\subsection{Setup}
We simulate Sec.~\ref{sec:model} with $M{\in}\{8,16,32,64\}$, $K{=}4$,
$\sigma^2{=}1$, equal targets $\bar{\gamma}_k{=}\bar{\gamma}$, and budgets
$P_m{=}P_{\rm base}{+}\delta_m$,
$\delta_m{\sim}\mathrm{Uniform}(-\Delta,\Delta)$,
$\Delta{=}0.3P_{\rm base}$, $P_{\rm base}{=}0.5$ ($P_0{\triangleq}MP_{\rm base}$).
We compare \textbf{RZF}~\eqref{eq:rzf_kkt}, \textbf{GIM-MMSE}~\eqref{eq:gim_final}
via Alg.~\ref{alg:gim} ($T_\Phi{=}8000$, $\eta_0{=}0.5$, $p{=}0.6$), the paired
per-trial \textbf{SOCP} (the PAPC-optimal peak over the GIM
family~\eqref{eq:minimax_phi}, our tightness benchmark), and
\textbf{WMMSE}~\cite{shi2011wmmse} as a non-peak-aware, sum-rate-optimal
reference---all scaled to targets, over $N_{\rm MC}{=}1500$ realisations ($700$
for Fig.~\ref{fig:fig3_gain_vs_M}). $95\%$ intervals are Wilson
(Fig.~\ref{fig:fig2_feasibility}) or bootstrap (Fig.~\ref{fig:fig3_gain_vs_M});
gains/gaps are paired dB means, excluding trials infeasible by common scaling
(mostly RZF at $M{=}8$).
\subsection{Peak Per-Antenna Power vs.\ SINR Target}
Figure~\ref{fig:fig1_power_vs_gamma} ($M{=}16$) plots $\max_m f_m(\bPhi^*)$ vs
$\bar{\gamma}$: GIM-MMSE runs \emph{parallel} to RZF with a \emph{constant}
${\sim}\NgainFone\,$dB reduction (the rescaling $c^2$~\eqref{eq:common_scale}
cancels in the ratio), tracking the paired SOCP to the ${\sim}0.7\%$ of
Sec.~\ref{sec:sims}-D.
\subsection{PAPC Feasibility Probability}
Figure~\ref{fig:fig2_feasibility} shows the infeasibility probability
versus imbalance $\Delta$ at $M{=}16$, $\bar{\gamma}{=}10\,$dB.
Even at $\Delta{=}0$, RZF violates PAPC in ${\sim}\NfeasZero\%$ of realisations;
GIM-MMSE removes essentially all violations to $20\%$ imbalance (${\leq}0.2\%$)
and stays low beyond (${\sim}\NfeasGIMhi\%$ at $50\%$),
matching Cor.~\ref{cor:feasibility}, while non-peak-aware WMMSE violates far more
(${\sim}57$--$82\%$).
\subsection{Scaling with Antenna Count}
Figure~\ref{fig:fig3_gain_vs_M} shows the gain over the baseline RZF
versus $M$ at $K{=}4$, $\bar{\gamma}{=}10\,$dB. The achieved gain rises
from $\NGAINlo\,$dB ($M{=}8$) to $\NGAINhi\,$dB ($M{=}64$) with clearly
diminishing returns, consistent with the logarithmic growth of the
baseline peak-to-average ratio (Prop.~\ref{prop:mrt}). WMMSE offers no
peak-power benefit over RZF (${-}1.0$ to ${+}0.05\,$dB across $M$).
Alg.~\ref{alg:gim} matches the paired SOCP optimum to within ${\sim}\NsocpM\,$dB
(${\sim}0.7\%$ mean, ${\le}2.03\%$ worst trial; $M{\le}64$).
\subsection{Verification and Channel Robustness}
\emph{Theorem checks.} Lemma~\ref{lem:sep} holds to machine precision (GIM vs
baseline-RZF SINRs agree to a relative $2{\times}10^{-15}$);
Thm.~\ref{thm:equalization} by the subgradient reaching the global SOCP optimum
and equalising the active antennas; Cor.~\ref{cor:feasibility} by the gain
growing with $M{-}K$; and Prop.~\ref{prop:mrt} by the $\Theta(\log M)$ peak growth
(Fig.~\ref{fig:fig3_gain_vs_M}).

\emph{Beyond i.i.d.\ Rayleigh.} The separation, convexity and existence results
are \emph{deterministic} (any full-rank $\Hmat$; the a.s.\ parts, any continuous
law). Re-running Alg.~\ref{alg:gim} under correlated Rayleigh ($\rho{=}0.6$) and
strong-LoS Rician channels ($\kappa{=}10$, random AoDs), the gain persists:
$2.8$/$3.3$\,dB at $M{=}16$/$32$ under correlation (as i.i.d.) and $2.3$\,dB under
LoS (smaller---steering vectors load antennas more evenly), still tracking SOCP
to ${<}1\%$. The feasibility edge \emph{widens}
under correlation: at $\Delta{=}0$ the baseline RZF is infeasible on $63\%$ of
channels vs GIM's $2.5\%$ ($200$--$400$ trials/point).
\section{Conclusion}
\label{sec:conclusion}
For MMSE/RZF precoding under PAPC the pseudo-inverse is \emph{not} the optimal
generalised inverse: a null-space correction reshapes per-antenna power at zero
SINR cost (Lemma~\ref{lem:sep}), strictly improving the peak a.s.\
(Prop.~\ref{prop:rzf_papc}, Thm.~\ref{thm:equalization}) and the PAPC feasibility
(Cor.~\ref{cor:feasibility}). A single-loop solver attains $\NGAINlo$--$\NGAINhi\,$dB
over the baseline RZF, within ${\sim}0.7\%$ of the SOCP optimum and robust across
correlated and LoS channels.
\let\oldthebibliography\thebibliography
\renewcommand{\thebibliography}[1]{%
  \oldthebibliography{#1}%
  \setlength{\itemsep}{-1.5pt}%
  \setlength{\parsep}{0pt}%
  \setlength{\topsep}{0pt}}
\bibliographystyle{IEEEtran}
\bibliography{GIM_references}

@article{wiesel2008,
  author    = {A. Wiesel and Y. C. Eldar and S. Shamai},
  title     = {Zero-Forcing Precoding and Generalized Inverses},
  journal   = {IEEE Trans.\ Signal Process.},
  volume    = {56},
  number    = {9},
  pages     = {4409--4418},
  year      = {2008},
  doi       = {10.1109/TSP.2008.924638}
}

@article{bjornson2014optimal,
  author    = {E. Bj{\"o}rnson and M. Bengtsson and B. Ottersten},
  title     = {Optimal Multiuser Transmit Beamforming: {A} Difficult Problem
               with a Simple Solution Structure},
  journal   = {IEEE Signal Process.\ Mag.},
  volume    = {31},
  number    = {4},
  pages     = {142--148},
  year      = {2014},
  doi       = {10.1109/MSP.2014.2312183}
}

@article{shi2011wmmse,
  author    = {Q. Shi and M. Razaviyayn and Z.-Q. Luo and C. He},
  title     = {An Iteratively Weighted {MMSE} Approach to Distributed
               Sum-Utility Maximization for a {MIMO} Interfering Broadcast
               Channel},
  journal   = {IEEE Trans.\ Signal Process.},
  volume    = {59},
  number    = {9},
  pages     = {4331--4340},
  year      = {2011},
  doi       = {10.1109/TSP.2011.2147784}
}

@article{yu2007transmitter,
  author    = {W. Yu and T. Lan},
  title     = {Transmitter Optimization for the Multi-Antenna Downlink with
               Per-Antenna Power Constraints},
  journal   = {IEEE Trans.\ Signal Process.},
  volume    = {55},
  number    = {6},
  pages     = {2646--2660},
  year      = {2007},
  doi       = {10.1109/TSP.2006.887156}
}

@article{bjornson2017massive,
  author    = {E. Bj{\"o}rnson and J. Hoydis and L. Sanguinetti},
  title     = {Massive {MIMO} Networks: Spectral, Energy, and Hardware
               Efficiency},
  journal   = {Foundations and Trends in Signal Processing},
  volume    = {11},
  number    = {3-4},
  pages     = {154--655},
  year      = {2017},
  publisher = {Now Publishers},
  doi       = {10.1561/2000000093}
}

@article{ahmed2018survey,
  author    = {Ahmed, Irfan and Khammari, Hedi and Shahid, Attique and
               Musa, Ahmed and Kim, Kwang Soon and De Poorter, Eli and
               Moerman, Ingrid},
  title     = {A Survey on Hybrid Beamforming Techniques in {5G}:
               Architecture and System Model Perspectives},
  journal   = {IEEE Commun.\ Surv.\ Tut.},
  volume    = {20},
  number    = {4},
  pages     = {3060--3097},
  year      = {2018}
}

@article{bjornson2020scalable,
  author    = {Bj\"{o}rnson, Emil and Sanguinetti, Luca},
  title     = {Scalable Cell-Free Massive {MIMO} Systems},
  journal   = {IEEE Trans.\ Commun.},
  volume    = {68},
  number    = {7},
  pages     = {4247--4261},
  year      = {2020}
}

@book{boyd2004convex,
  author    = {S.~Boyd and L.~Vandenberghe},
  title     = {Convex Optimization},
  publisher = {Cambridge University Press},
  address   = {Cambridge, UK},
  year      = {2004}
}

@article{hjorungnes2007complex,
  author  = {Hj{\o}rungnes, Are and Gesbert, David},
  title   = {Complex-Valued Matrix Differentiation:
             Techniques and Key Results},
  journal = {IEEE Trans.\ Signal Process.},
  volume  = {55},
  number  = {6},
  pages   = {2740--2746},
  month   = jun,
  year    = {2007},
  doi     = {10.1109/TSP.2007.893762}
}

@article{zhang2025ris,
  author  = {X. Zhang and D. Xu and J. Wang and C. Jiang and S. Song and M. {Di Renzo}},
  title   = {{RIS}-Aided Secure Communications With Regularized Zero-Forcing Precoding},
  journal = {IEEE Trans.\ Inf.\ Forensics Security},
  volume  = {20},
  pages   = {5843--5858},
  year    = {2025},
  doi     = {10.1109/TIFS.2025.3575624}
}

@article{kaviani2013optimal,
  author  = {Saeed Kaviani and Witold A. Krzymie{\'n}},
  title   = {Optimal Multiuser Zero-Forcing with Per-Antenna Power Constraints for Network {MIMO} Coordination},
  journal = {EURASIP Journal on Wireless Communications and Networking},
  volume  = {2011},
  number  = {1},
  pages   = {190461},
  year    = {2011},
  doi     = {10.1155/2011/190461}
}

@inproceedings{suzuki2018papr,
  author    = {Mikihito Suzuki and Yoshihisa Kishiyama and Kenichi Higuchi},
  title     = {Combination of Beamforming with Per-Antenna Power Constraint and Adaptive {PAPR} Reduction Method Using Null Space in {MIMO} Channel for Multiuser Massive {MIMO-OFDM} Transmission},
  booktitle = {Proceedings of the 21st International Symposium on Wireless Personal Multimedia Communications},
  year      = {2018}
}

@misc{boyd2003subgradient,
  author       = {S.~Boyd and J.~Park},
  title        = {Subgradient Methods},
  howpublished = {Lecture notes, Stanford Univ. EE364b},
  year         = {2014}
}

@article{nedic2001incremental,
  author  = {A.~Nedi\'{c} and D.~P.~Bertsekas},
  title   = {Incremental Subgradient Methods for Nondifferentiable Optimization},
  journal = {SIAM J. Optim.},
  volume  = {12},
  number  = {1},
  pages   = {109--138},
  year    = {2001},
  doi     = {10.1137/S1052623499362111}
}

@article{aghdam2021lowspatial,
  author  = {Sina Rezaei Aghdam and Thomas Eriksson},
  title   = {Low Spatial Peak-to-Average Power Ratio Transmission for Improved Energy Efficiency in Massive {MIMO} Systems},
  journal = {Sensors},
  volume  = {21},
  number  = {16},
  pages   = {5534},
  year    = {2021}
}
\end{document}